\documentstyle[psbox]{WORKSHOP} 

\input{psfig.sty}

\def\etal{{\it et al.~}}
\def\cm2-sec{\ cm$^{2}$-sec}
\def\la{\hbox{\rlap{$<$}\lower.5ex\hbox{$\sim$}\ }}
\def\ga{\hbox{\rlap{$>$}\lower.5ex\hbox{$\sim$}\ }}
\def\deg{$^{\circ}$~}
\def\arcmin{$\,^\prime$~}

\begin{document}  
  
 
\title{Proposed (to) EXIST: Hard X-ray Imaging All Sky Survey/Monitor}  
 

\author{
J.E. Grindlay,$^1$ T.A. Prince,$^2$ F. Harrison, $^2$ 
N. Gehrels, $^3$ C.J. Hailey, $^4$ \\ 
B. Ramsey,$^5$ M.C.  Weisskopf,$^5$ G.K. Skinner,$^6$ P. Ubertini,$^7$\\
\\[12pt]  
%
$^1$Harvard-Smithsonian Center for Astrophysics, Cambridge, MA 02138\\ 
$^2$California Institute of Technology, Pasadena, CA 91125\\
$^3$Goddard Space Flight Center, Greenbelt, MD 20771\\
$^4$Columbia University, New York, NY 10027\\
$^5$Marshall Space Flight Center, Huntsville, AL 35812\\
$^6$University of Birmingham, Birmingham B15 2TT, UK\\
$^7$IAS, I-00044 Frascati, IT \\
{\it E-mail: josh@cfa.harvard.edu} 
}

\abst{  
 The hard x-ray (10-600 keV) sky is inherently time variable  
and yet has neither been surveyed nor been monitored with  
a sensitive imaging telescope. The Energetic X-ray Imaging  
Survey Telescope (EXIST) is a mission concept, proposed   
for MIDEX, which would conduct the first imaging all-sky  
hard x-ray survey as well as provide a sensitive all sky monitor  
(ASM). With  $\sim$60\% sky coverage each orbit, and full sky 
coverage each $\sim$50 days,  hard x-ray studies of  
gamma-ray bursts, AGN,  
galactic transients, x-ray binaries 
and accretion-powered pulsars can be 
conducted over a wide range of timescales. 
We summarize the scientific objectives of EXIST for both the  
survey and monitoring objectives. We describe the mission  
concept and the instrumentation approach, which incorporates  
a large area array of  Cd-Zn-Te (CZT) detectors, as well as  
some of our ongoing development of CZT array detectors. 
} 
 
\kword{surveys: X-rays --- X-ray sources: binaries, black holes, AGN 
--- GRBs --- instruments: all sky monitors} 
 
\maketitle

\section{Introduction} 
  
The Energetic X-ray Imaging Survey Telescope (EXIST) was proposed (in  
December 1994) and accepted (April 1995) as one of  the 27  
New Mission Concept (NMC) studies  for   
a satellite-borne astrophysics mission. It would conduct the first  
imaging survey of the sky at hard x-ray energies (10-600 keV) with a  
sensitivity some 100$\times$ greater than the only previous all-sky  
survey carried out by HEAO-A4 experiment in 1978-80   
(Levine et al 1984). The need, and priority, for such
an all sky imaging hard x-ray survey mission has been pointed out 
in the recent report of the NASA Gamma-Ray Program Working Group. 
An overall description of the initial EXIST   
concept is given by Grindlay et al (1995) (and on the Web site   
{\it http://hea-www.harvard.edu/EXIST/EXIST.html}).   
EXIST has been developed  
extensively in the course of preparation and submission of  
a successful ``Step 1" and invited (1995) ``Step 2''   
proposal for the MIDEX program. The mission is being 
further developed for the next MIDEX solicitation, 
including additional laboratory development and balloon 
flight testing of the Cd-Zn-Te detectors proposed, and community 
input for the mission design has been sought. A detailed summary and 
description of the Mission Concept Study will be given by     
Grindlay et al (1997).   

\section{Overview of EXIST} 

%
%
\begin{figure*}[t]  
\centering  
\vspace*{-0.7in}
\hspace*{0.3in}
\psfig{figure=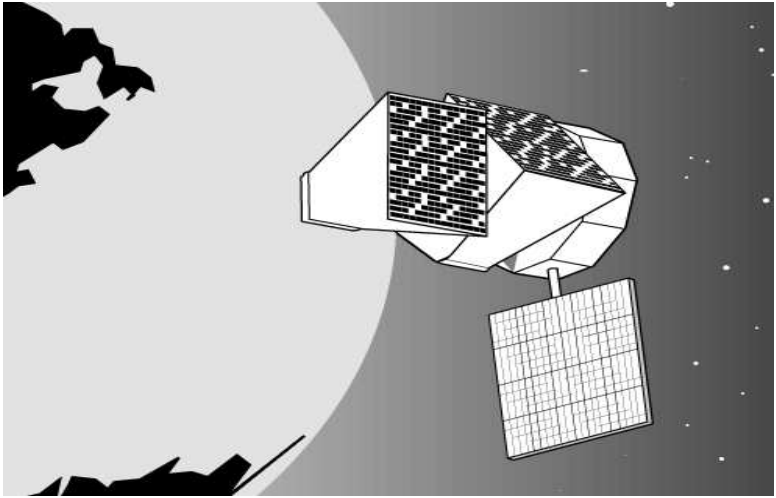,height=4.5in,width=4.5in,angle=0}
\vspace*{-0.7in}
\caption{Schematic view of EXIST in orbit showing layout of 
coded aperture telescopes and spacecraft.}  
\end{figure*}

A survey telescope at hard x-ray energies can   
be constructed as a coded aperture telescope with a field of   
view of up to $\sim$45\deg without significant  projection   
effects or collimation by the mask aperture, assumed planar.   
Such a telescope   
would execute a continuous scan rather than fixed target pointing   
to cover a maximum sky fraction in minimum time and would   
be more sensitive (above 30 keV) than a scanning 
grazing incidence (multi-layer)   
telescope of comparable (or even larger) physical size though smaller   
effective area and field of view. For example, 
a multi-layer telescope with (currently ambitious) parameters of   
FOV $\sim$ 10\arcmin and effective area A$_{eff} \sim$ 500cm$^2$   
in a scanning (ROSAT-like) satellite mission would be a factor of   
$\sim$10 {\it less} sensitive than a wide-field (FOV $\sim$ 20-40\deg)   
scanning coded aperture telescope with A$_{eff} \sim$ 5000 cm$^2$   
in the 30-100 keV band. 
In addition, the wide-field coded aperture imager allows the   
survey to extend up to the poorly explored 100-600 keV range,  
totally unaccessible to focussing optics (except for Bragg  
concentrators, which can work only in a very narrow energy band).   
    
EXIST would incorporate two coded aperture telescopes, each 
with FOV = 40\deg $\times$ 40\deg (above $\sim$40 keV; at 10-30 keV 
a low energy collimator would restrict the FOV to 3.5\deg in one 
dimension), for a  combined FOV =  40\deg $\times$ 80\deg,.
A Cd-Zn-Te (CZT) detector array of total area of 2500 cm$^2$ is 
at the focal plane of each telescope.   
The schematic layout of the two telescopes and 
a possible spacecraft implementation is shown in Figure 1.   
  
Because of its very large FOV and large area detectors with   
high intrinsic resolution (both spatial and spectral), EXIST 
could approach the unprecedented all-sky sensitivity  
shown in  Figure 2.   
The sensitivity plots are for EXIST for its proposed   
9-month all-sky survey (followed by a pointed mission   
phase), which would allow total integration times of   
$\sim$10$^6$ sec for any source.

Although the mission is designed primarily as a survey 
and monitor mission, 
it could be operated as a pointed (observatory) 
mission for selected high priority 
targets and very long exposure times (e.g. M31 GRB
 and BHC surveys; galactic 
bulge survey) in which case even greater total 
exposure times and thus sensitivities 
can be achieved.
 
\section{Scientific Objectives: All Sky Survey and Monitoring} 
 
The scientific objectives for  the EXIST mission may be  
summarized either by classes of object to be studied or  
by the broad objectives of surveys and monitoring studies.  
In this paper we emphasize  the  
monitoring and general time-variability studies that  
EXIST can carry out. However, since many of these  
require the hard x-ray samples and populations of  
(many) classes of object to be first studied and established  
beyond the currently limited samples, we begin with the  
survey objectives. 

%
%
\begin{figure*}[t]  
\centering  
 
\hspace*{-3.2in}  
\psfig{figure=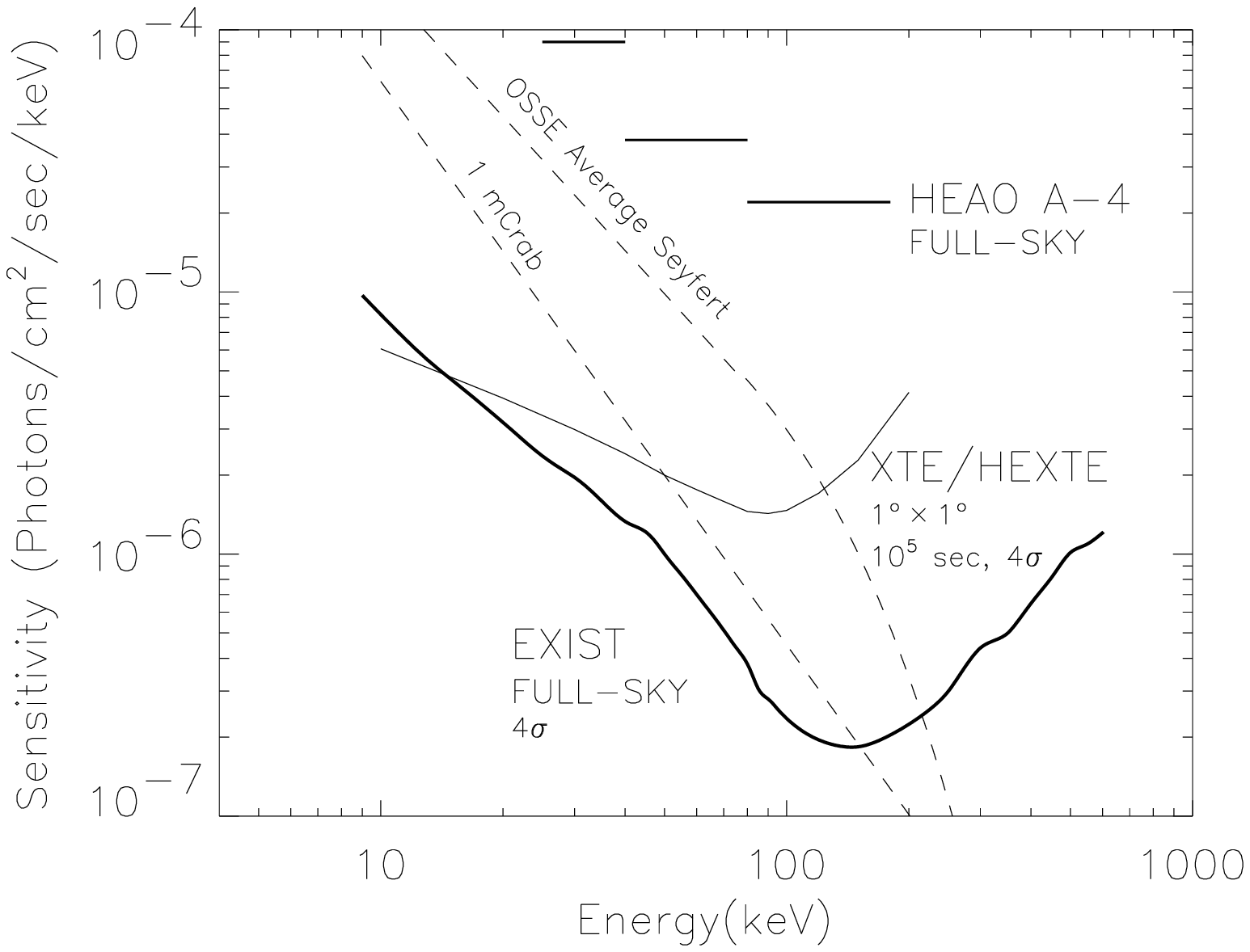,height= 3.2in,width=3.9in,angle=0}  
  
\vspace*{-3.2in}  
  
\hspace*{3.35in}  
\psfig{figure=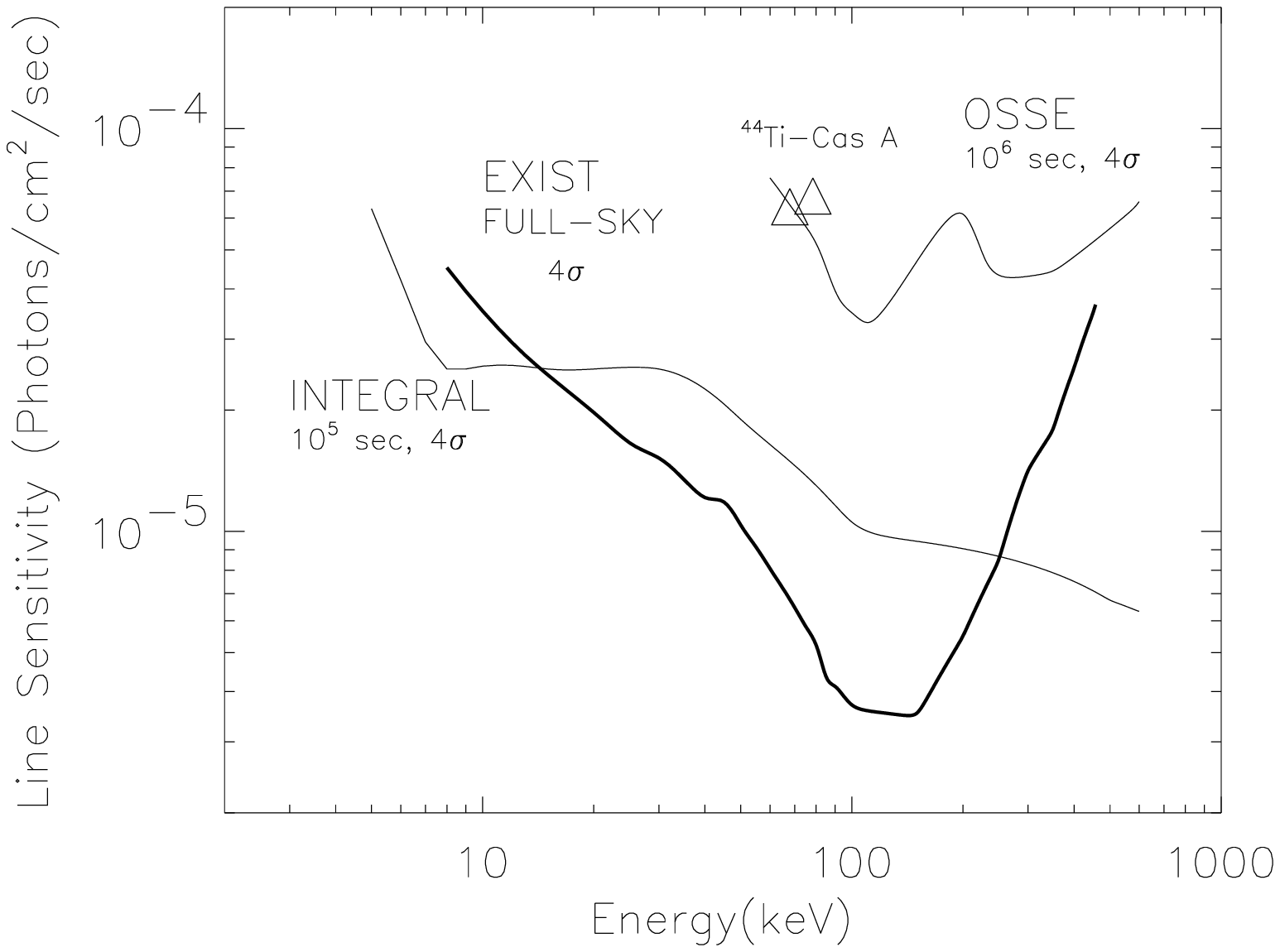,height= 3.2in,width=3.9in,angle=0}  
  
\caption{Sensitivity of EXIST (MIDEX) for continuum (left)  
and narrow line (right) spectra and compared with other missions.}  
\end{figure*}

\subsection{Surveys} 
 
{\it Hard x-ray spectra and luminosity function of AGNs:} Active galactic  
nuclei (AGN) are now measured by OSSE to have hard spectra with breaks  
typically in the 50-100 keV range for Seyferts (cf. Zdjiarski et al 1995)  
and with multi-component or non-thermal spectra extending to  
higher energies likely for the  
blazars. EXIST will have an all-sky sensitivity 
some 10$\times$ better than  
that needed to detect the ``typical'' Seyferts seen with OSSE.  
More than 1000 AGN should be detected in the all sky survey and   
EXIST has the required sensitivity in   
the poorly explored 30-200 keV band to measure 
accurate spectra for all known AGN detected   
with Ginga or with the Einstein slew survey.  
 
A major objective for AGN studies and surveys is the 
detection and inventory of  heavily obscured or self-absorbed AGN.  
Such objects, primarily Seyfert 2's but also 
including (some) star-formation  
galaxies, are now being discovered in pointed observations with SAX  
(cf. Piro, these proceedings) and will also likely 
be discovered with the focussing  
ABRIXAS all sky survey up to $\sim$10 keV(cf. Staubert, these  
proceedings). However,  the most heavily obscured objects, with absorbing  
column densities \ga10$^{24-25}$cm$^{-2}$ yielding low energy cutoffs  
in the 5-10 keV range, will be more readily detected with EXIST. This,  
together with the EXIST measure of the hard 
spectra and thus total luminosity  
of  the still larger sample of obscured AGN with lower cutoffs  
discovered with comparable sensitivity ($\sim$0.1mCrab) with  
ABRIXAS, will yield the first measure of the luminosity function of  
this important yet poorly studied class of AGN.  
 
{\it Survey for black hole and neutron star compact binaries.}  
Studies of compact objects over a wide range of   
timescales and luminosity are possible 
throughout the Galaxy. A deep galactic survey  
for x-ray binaries containing  black holes vs. neutron 
stars and pulsars will allow the  
relative populations of black holes in the Galaxy to be constrained.   
All previous galactic hard x-ray surveys have been constrained to 
the brightest decade in source flux (and luminosity); EXIST will 
extend this 1-2 decades deeper. Whereas the INTEGRAL galactic 
plane survey(s) will also make great strides, the EXIST (all sky) survey 
will be more sensitive and not be limited to the central $\sim\pm$7\deg 
of galactic latitude covered by the smaller FOV of INTEGRAL. The EXIST
survey will also sample the galactic plane on a wider range of
timescales. 
 
{\it Emission line surveys: hidden supernovae via $^{44}$Ti  
emission and 511 keV sources:}  
The array of CZT imaging detectors proposed for EXIST  
achieves high spectral resolution (e.g. $\sim$5\% at 60 keV).   
Thus emission line surveys can be conducted.   
The decay of $^{44}$Ti (lines at 68 and 78 keV) with long (68 y)  
halflife allows a search for the long-sought population of  
obscured supernovae in the galactic plane at sensitivities   
significantly better than the possible detection of Cas-A (cf. Figure  
2). These objects would likely appear as discrete (unresolved)   
emission line sources. Similarly, 511 keV emission from black hole   
binaries (or AGN) can be searched for (e.g. in transient outbursts),    
and the diffuse galactic 511 keV emission imaged with sensitivity   
comparable to OSSE (cf. Figure 2).   
  
{\it Study of the diffuse hard x-ray background:} 
The spectra of a significant sample of AGN will test the AGN   
origin of the diffuse background  for the poorly  
explored hard x-ray band. Because the background measured by  
the EXIST detectors below 100 keV is dominated by the cosmic diffuse  
spectrum, its isotropy and fluctuation spectrum can be studied   
with much higher sensitivity than before.

\subsection{Monitoring} 

EXIST surveys 60\% of the sky each day, yielding $\sim$10 orbits/day 
(allowing for SAA, etc.) 
$\times$ \ga10 min exposure/orbit or \ga6000 sec/day for each source 
observed. This yields a daily flux sensitivity (30-100 keV) 
of $\sim$1-2 mCrab, sufficient 
for the brightest AGN and essentially all known accretion-powered 
binaries in the Galaxy. Pulsar timing allows even fainter flux limits, 
as demonstrated with the extensive BATSE 
monitoring project (cf. Bildsten et 
al 1997). Over one sky survey epoch ($\sim$50d), each source is observed 
for \ga25d, and variability flux limits for $\sim$month timescales are 
thus $\sim$0.3mCrab. 

{\it Faint hard x-ray transients: black hole population in Galaxy:} 
The sensitivity to $\sim$1-10d transients is \ga30 $\times$ better than 
BATSE, so that the low resolution occulatation-imaging 
survey for faint transients being conducted with BATSE (cf. 
Grindlay et al 1996) can be extended to correspondingly lower 
outburst luminosities or greater source distances. With a 
1-10d sensitivity of only $\sim$1 mCrab, BH transients  can be 
detected with their characteristic peak luminosities of 
$\sim10^{37-38}$erg/s (10-100 keV) out to 100 kpc, so that 
the LMC/SMC can be surveyed in hard x-rays for the first time. 
Since the transients containing BHs are characterized by 
hard x-ray spectral components (extending usually to \ga100 keV) 
which are both more luminous (at peak) and longer lived than 
than the hard tails found for NS systems, the hard x-ray 
all-sky survey provides perhaps the best strategy for 
discovering x-ray binaries containing BHs. 

{\it Monitoring and Study of X-ray Pulsars:}  
The measurement and monitoring of spin periods, pulse 
shapes and luminosity/spectra   
of a large sample of accretion-powered pulsars would 
extend the BATSE sample  of Bildsten et al (1997)  
to the entire sample (\ga30) of known   
accretion-powered pulsars. The high spectral resolution 
of the CZT detectors on EXIST will allow high sensitivity 
studies of cyclotron features in pulsar spectra, greatly 
extending current RXTE/HEXTE studies of relatively few 
objects to a much larger sample. 
 
{\it Studies of  Gamma-ray Bursts:}   
EXIST would have a GRB sensitivity approximately 20$\times$ that of  
BATSE so that a 2 month pointed exposure could   
both detect and map a halo in the Andromeda  
galaxy (M31) should one exist as a component of 
the GRB population. Given the  
observed GRB logN-logS relation, EXIST should detect GRBs overall  
at about 1/2 the rate, or $\sim$0.5/day, as BATSE with its   
much larger FOV but reduced sensitivity. GRBs will be  
located to \la1-5\arcmin positions, thereby providing definitive tests  
of repeaters. Bright burst positions and spectra could be brought down  
in real time for automated followup searches.

\section{Detector, Telescope and Overall Mission Concepts} 
  
\subsection{Detector Concept}  
  
The sensitivity and resolution (both angular and spectral) desired for  
EXIST can be achieved with a large area array of pixellated CZT detectors  
used as the position-sensitive readout for a wide-field coded aperture  
telescope. The wide-field needed for the monitoring as well as survey  
objectives, combined with the desired sensitivity up to $\sim$600 keV,  
lead to a relatively thick (3-5mm, minimum) detector with moderately  
large pixel size. For the 40\deg FOV, and a 5mm thick detector, the  
pixels need be \ga5mm $\times$ tan(20\deg) $\sim$ 1.9mm across for  
minimal charge spreading due to projection effects.  
Since CZT crystals are much more readily available (currently, at least)  
in \la10-12mm crystal sizes, this leads 
naturally to a unit detector element   
of 12 mm (square) $\times$ 5mm (thick) on which a 4 $\times$ 4  
array of  2.5mm (square) pixels is inscribed with 0.5mm gaps between  
pixels. Such a detector pixel size/thickness ratio should yield a  
reasonable small pixel effect advantage 
(cf. Barret et al 1995), and studies  
are currently underway by us (at CfA) to verify this.  
  
The  individual 4 $\times$ 4 pixel detector elements 
would be grouped into a 2 $\times$ 2 sub-array for 
a 64-channel basic detector element (BDE) read out by   
a preamp-shaper-multiplexer ASIC readout circuit with very low  
power dissipation) \la1mW/channel). The ASIC is self-triggered  
and multiplexes its 64-channel output to a 
processing chain which could encode the 1, 4, or   
16 peak channels (configured on command) so that   
multi-site detection (e.g. Compton events and 
internal background rejection)   
could be accomplished. We are currently investigating packaging  
schemes to assemble these unit detector 
elements (BDE) into a tiled array.  
It is likely that tiling would be done by combining a   
sub-array of 4 $\times$ 4 BDEs  contiguously into a single  
basic detector module (BDM), with area $\sim$100 cm$^2$ (depending  
on final choice of unit detector and thus also pixel size).  
 
The 40\deg collimator could be either passive (e.g. Ta slats) 
or, more effective, active (BGO). We are studying the 
tradeoffs but an active collimator would allow each BDM 
to be separately shielded with a rear (2cm)/side (0.5-1cm) BGO shield
for a well-type geometry yielding both collimation and lower 
background (due to forward-hemisphere rejection of internal CZT 
activation background as well as lack of production in the passive 
collimator). 
For the 40\deg (FWHM) field of view desired for each EXIST telescope,   
the BGO shield/collimator blade height is 12cm so that the 
BDM modules are nearly cubic. 
Each of the two complete EXIST detectors would thus consist of   
an array of 5 $\times$ 5 or   
25 such BDMs. These modules would be close-packed   
and mounted in a common frame. Although the   
collimator shields will require a net gap of $\sim$1.5 cm between   
each of the 5 BDMs across the detector array, this does not affect   
image reconstruction but only makes the detector $\sim$8 cm larger on a side.   
  
\subsection{Telescope Concept}  
  
The coded aperture telescopes  
can be relatively compact design with coded mask at 
focal length 1.4m and mask  
pixel size 5mm. This yields an imaging resolution of 12\arcmin,   
which is appropriate to resolve even the most crowded galactic bulge   
fields at the high sensitivity expected. In order to cover the   
full 40\deg field of view, each telescope   
would have a URA mask of approximate dimensions   
1.2m $\times$ 1.2m and format 257 $\times$ 255 to fully image the 40\deg   
FWHM field of view. However, since the BGO collimator on each BDM   
segment of the detector array would produce partial coding for   
sources off-axis, the mask must be either smaller format and repeated   
(e.g. 4 contiguous 129 $\times$ 127 masks, 
leading to ambiguous source positions) or random.   
A random mask would be, in any case, as effective as a URA   
of such large format.   Since   
the coded mask should not collimate the image significantly,   
the mask thickness is   
restricted to be \la5mm, which (for Ta mask elements) restricts its   
upper energy limit to be \la600 keV for partial shadowing.
  
\subsection{Mission Concept}

EXIST is proposed as a MIDEX with nominal lifetime 
of 2 years (although with no consumables, an extended mission 
is desirable). The baseline plan is for 
the all sky survey to be conducted in the first 10 months (including 
a 1 month verification phase) which allows 6 passes through the sky. 
The entire sky is then effectively surveyed to a total exposure 
of $\sim1.5 \times 10^6$ sec for any source. 
The two-telescope combined FOV of 80\deg 
is pointed North-South, and the pointing direction ``nods'' toward the
poles during the high latitude portion of the orbit so that all-sky 
coverage is uniform over the survey to within \la10\%. 
A pointed mission phase of 2 months within the first year would allow 
the deep survey of M31 (for GRBs and BHCs) 
and the galactic center and/or 
the LMC/SMC to be conducted (during M31 occultations). These deep
pointed surveys would more than double the total exposure on these 
targets that has been acquired during the all sky survey phase, as
well as allow for broader timescales of coverage. The second year of 
the nominal mission would be a pointed phase, with a series of 
deep pointings ($\sim$1 month each) designed to both allow deep surveys 
and TOOs as well as to effectively extend the all sky survey and 
ASM function given the large FOV covered. Data 
from the entire mission is open (by proposal) 
to the entire community, and pointing directions/durations 
(and surveys) are proposed by a GI program.

\section{Ongoing CZT and Array Studies}

As part of the effort to both conduct the EXIST Mission Concept study   
and optimize the design for a future MIDEX proposal as well   
as a prototype balloon-borne implementation, we are   
conducting a variety of studies of CZT detectors and array technologies   
at our respective institutions. Here we outline briefly the projects 
underway at CfA; space does not permit description of 
the significant development efforts underway at the other 
institutions in the EXIST collaboration. 
 
{\it Balloon Flight Tests of Backgrounds and Shielding Efficiency:}  
The large neutron cross section(s) for Cd, which result in   
prompt gamma-ray decays, may yield high internal 
backgrounds for CZT detectors in space. Balloon flight tests of single   
isolated CZT detectors by the GSFC and Caltech groups in   
May and September-October, 1995, suggested disturbingly   
large in-flight backgrounds compared to those expected for   
similar scintillation detectors (e.g. Parsons et al 1996). However   
the GSFC measurement of a marked reduction in background with   
an external anti-coincidence shield (NaI) suggested this could   
be effectively reduced by suitable active shielding.   
  
In collaboration with Caltech and JPL, we assembled   
a flight unit to test the prompt anti-coincidence shielding   
efficiency of a planar BGO shield immediately behind the   
CZT detector plane, as proposed originally for EXIST (a 
future experiment will test the possible well-type geometry). The   
BGO (75mm diameter $\times$ 75mm thick,   
and supplied by JPL) was centered below a single element CZT detector   
(10mm $\times$ 10mm $\times$ 2mm, and supplied by eV Products   
to Caltech). The detector-shield and preamp was mounted in a   
pressure vessel and shielded with a 1.8mm thick Pb + 0.8mm thick Sn   
and 1.2mm Cu graded shield to simulate the approximate grammage   
of the passive high energy collimator in the baseline EXIST detector. 
The raw CZT and BGO (shield) detector   
preamp outputs are interfaced to shaping amps and digital (discriminator   
and 12 bit ADC) electronics built at CfA to interface to the   
flight computer and data system   
for the EXITE2 balloon-borne telescope (Lum et al 1994, 1997). 

A balloon flight was (finally) obtained on May 8, 1997. 
The background spectrum in  
the CZT is reduced by a factor 
(energy-dependent) of $\sim$4-8 with the BGO veto and  
is at flux level at 100 keV of $\sim1.0 \times 10^{-3}$ cts/\cm2-sec, 
or within a factor  
of 2 of the EXITE2 phoswich background level measured simultaneously.  
With the additional shielding efficiency possible with a well-type active  
BGO collimator, the background may be reduced another factor of  $\sim$2.  
Full details are given in Bloser et al (1997).

{\it Balloon Flight Measure of Neutron Backgrounds:}  
In order to fully calibrate the CZT background and shielding   
experiment so that balloon results may be extrapolated to   
the full space environment, a simultaneous measure of the neutron flux   
experienced by the detector is desirable. The atmospheric neutron   
fluxes  as tabulated by Armstrong et al (1973) are   
sufficiently uncertain (probably by a factor of \ga2) that   
we have attempted to measure the flux by a simple passive experiment:   
an array (7 $\times$ 6) of gold foils (each $\sim$6cm$^2$) mounted   
on top of the gondola in which the n-$\gamma$ reaction

Au-197(n,$\gamma$)~~$\rightarrow$~~~Au-198(,e)Hg-198

\noindent  
was (attempted to be) measured after the 
flight by observing the resulting   
412 keV decay $\gamma$-ray (2.7d halflife) with a low   
background Ge spectrometer at JPL (by L. Varnell).  This   
experiment, conducted in collaboration with G. Skinner   
and L. Varnell, is currently being analyzed.

{\it Spatial Uniformity of CZT:}  
Pixellated CZT detector arrays, as proposed for   
EXIST, will require   
relatively uniform response across both the projected surface area   
and depth of the detector elements. Non-uniformities of detector   
response can be calibrated out (by flat fielding) but will be simplified   
to the extent the detectors are uniform (and may be less of a   
problem with the relatively large pixel detectors for EXIST   
than with small pixel CZT imagers for focussing optics).   
We have conducted a program of  mapping the spectral   
response of single detectors and comparing the observed   
variations with IR micrographs (obtained at eV Products)   
of the detector to correlate   
spectral response with grain boundaries and inclusions in the   
detector. Spectra (Am-241) obtained in a 3 $\times$ 3 raster   
scan of a 0.5mm beam across  a 
4mm $\times$ 4mm $\times$ 3mm CZT detector   
show variation in spectral response which correlates  
with the grain boundaries    
as well as inclusions and precipitates. Results were presented at 
the NASA-SEUS Workshop in December 1996.
  
{\it Development of PIN Readouts for CZT Detectors:}  
CZT detectors are conventionally fabricated with metal (gold)   
contacts deposited directly on the CZT crystal. These   
metal-semiconductor-metal (M-S-M) detectors are of course the subject   
of intense development and are baselined for EXIST. However, they   
can suffer from limitations of charge collection efficiency (though   
at least partially overcome with ``small pixel'' electrodes; cf.   
Barrett et al 1995) and poor ohmic contacts. Several groups, most   
recently SBRC (Hamilton et al 1996) have investigated alternative   
readouts incorporating P-I-N junctions. The Spire Corp. (Bedford, MA)   
has developed a  new method for fabrication of P-I-N electrodes on   
CZT by using CdS(p-type) and ZnTe(n-type) layers deposited by thermal   
evaporation on both high pressure Bridgman (HPB) CZT crystals (from eV  
Products) as well as lower cost vertical Bridgman (VB) crystals (from   
Cleveland Crystals), and the results appear very encouraging.   
A P-I-N configuration for CZT offers the 
possibilities of both further enhanced energy resolution (due 
to the higher bias voltage possible) and/or lower detector element 
cost (due to lower resistivity CZT being possible). 
At CfA we are testing these P-I-N readout CZT detectors which offer   
advantages of improved charge collection and ease of fabrication for   
their use as thick detectors. We are working with Spire to fabricate   
a 4 $\times$ 4 array P-I-N detector and 
comparison M-S-M array detector on 10mm $\times$ 10mm   
$\times$ 5mm CZT substrates for laboratory evaluation and 
balloon flight tests of background and   
uniformity of response. 
  
{\it Development of Thick CZT Detector Array Readouts:}  
Thick detectors (5mm or greater) as desired for EXIST pose   
special challenges for the optimum design of the detector and   
readout. In particular, the electric field configuration   
needed for the small pixel effect (Barrett et al 1995) must be   
carefully considered, and the effects of charge diffusion and   
spreading become more important. We are exploring these effects   
in collaboration with both Spire Corp. and 
the RMD Corporation (Watertown, MA), who   
have just completed fabrication   
of a prototype 4 $\times$ 4 array M-S-M detector 
on a 10mm $\times$ 10mm   
$\times$ 5mm CZT substrate.  Initial results appear very   
promising and will be reported in Shah et al (1997).   
At CfA, we are now testing this array for its small-pixel effect   
properties and we have developed an interface to a 16-channel 
ASIC preamp/shaper supplied by the IDE Corp (Oslo). 
The CZT array detectors 
(10mm; 16 channels) are mounted on chip carriers for easy plug-in 
inter-comparison of the M-S-M vs. P-I-N detectors through the 
same ASIC readout system, with  
results  presented in a forthcoming paper by Bloser et al (1997).   
A balloon flight test of both array detectors, with a well-type 
BGO shield, may be conducted in 1998.

\bigskip   
 
This work was supported in part by NASA grant NAG8-1212.

\section*{References} 
 
\re 
Armstrong, T.W., Chandler, K.C. and Barish, J. 1973,  
JGR, 78 (16), 2715.  
 
\re 
Barrett, H., Eskin, J. and Barber, H 1995,  Phys. Rev. Letters,  75, 156.  
 
\re
Bildsten, L. \etal 1997, ApJ. Suppl., in press.

\re 
Bloser, P., Grindlay, J., Narita, T. \etal 1997, in preparation.  
 
\re 
Grindlay, J., Prince, T., Gehrels, N.., Tueller, J., Hailey, C.   
\etal 1995,  Proc. SPIE,  2518, 202.  

\re 
Grindlay, J., Prince, T., Gehrels, N.., \etal 1997, in preparation. 

\re 
Grindlay, J. Barret, D., Bloser, P. \etal 1996, A\&A, 120, 145. 

\re 
Hamilton, W., Rhiger, D., Sen, S., Kalisher, M., Chapman, G.   
and Millis, R. 1996, Jour. Elec. Mat., 25, 1286. 
 
\re  
Levine, A.M. \etal 1984,  ApJ Suppl., 54, 581. 
 
\re
Lum, K.S. \etal 1994, IEEE Trans. Nucl. Sci., NS-41, 1354.

\re
Lum, K.S. \etal 1997, NIM A, in press.

\re  
Parsons, A. \etal 1996, Proc. SPIE, 2806, 432.  
 
\re  
Paul, J. \etal 1991, Adv. Sp. Res.,  11 (8), 289.  
 
\re  
Shah, K., Cirignano, L., Klugerman, M., Dmitreyev, Y.,   
Grindlay, J. \etal 1997, IEEE Trans. Nucl. Sci., submitted.   
  
\re
Zdziarski, A., Johnson, W., Done, C. \etal 1995, ApJ, 438, 63.
 
\label{last} 
 
\end{document}